\documentclass[twocolumn,showpacs,assymb,amsmath]{revtex4}
\usepackage{mathrsfs}
\usepackage{txfonts}
\usepackage{graphicx}
\usepackage{amssymb}
\usepackage{amsmath}
\usepackage{color}

\begin{document}

\title{Hall conductance  of two-band systems in  a quantized field}

\author{Z. C. Shi$^{1,2}$, H. Z. Shen$^{1,2}$, and X. X. Yi$^1$
\footnote{Corresponding address: yixx@nenu.edu.cn}}
\affiliation{$^1$Center for Quantum Sciences and School of Physics,
Northeast Normal University, Changchun 130024, China\\
$^2$School of Physics and Optoelectronic Technology, Dalian
University of Technology, Dalian 116024, China }

\begin{abstract}
Kubo formula gives a linear response of a quantum system to external
fields, which are classical and weak with respect to the energy of
the system. In this work, we take the quantum nature of the external
field into account, and define a Hall conductance to characterize
the linear response of a two-band system to the quantized field. The
theory is then applied to topological insulators. Comparisons with
the traditional Hall conductance are presented and discussed.

\end{abstract}

 \maketitle

\section{Introduction}
The integer quantum Hall  effect(IQHE) is manifested   by a
remarkably precise quantization of the transverse conductance in
two-dimensional electron systems in presence of a strong
perpendicular magnetic field. Its discovery\cite{klitzing80,tsui82}
has had  profound implications for the understanding of  matter, and
it may find potential applications in quantum information
processing\cite{simon08}. The integer quantum Hall effects  can be
understood in the single particle framework\cite{hasan10,qi10}:
Charged particles in a magnetic field form Landau levels with energy
splitting that is proportional to the strength of the magnetic
field, and when an integer number of Landau levels are filled, the
Hall conductance is quantized and characterized by the TKNN
number\cite{thouless82} that is now treated  as a topological
invariant called ¡°Chern number¡±. This topological understanding of
the IQHE is a remarkable step of progress, opening up the field of
topological electronic states in condensed matter physics. Later,
Haldane\cite{haldane88} found that a periodic 2D honeycomb lattice
without net magnetic flux can in principle support a similar integer
quantum Hall effect. This result suggested that certain materials,
other than the 2D electron gas under magnetic field, can have
topologically non-trivial electronic band structures, which can be
characterized by a non-zero Chern number. Such materials are called
topological  insulators now.

In contrast to ordinary band insulators, topological
insulator\cite{konig07,roth09,qi11} comes with gapless chiral   edge
states that each carries a quantum of conductance, $\frac{e^2}{h}$.
The number of edge states is mathematically given by the value of
the topological invariant, namely the Chern number, that can only
assume integer values similar to winding numbers. The integer nature
of the Chern number is what makes the edge states, and hence the
quantization of the conductivity.

Physically, the quantized conductance can be derived by linear
response theory.  In the context of quantum statistics, the
exposition of the linear response theory can be found in the paper
by Ryogo Kubo\cite{kubo57}, which defines   particularly the Kubo
formula.  This formula gives a linear response of quantum systems to
external classical fields.  Particularly, it considers the response
to a classically electric filed  of an otherwise stationary
observable, say current. The goal for us in this work is to answer
the following question: When the field is quantized, how a quantum
system responses to that field?

The answer to this question is not trivial. Firstly, this answer
conceptually  contributes to the broader question of how quantum
systems respond to a quantized driving. A simple setting is provided
by a two-band model  (that can describe TIs) driven by a single mode
electromagnetic field with frequency $\omega$, with the Hall current
denoting the response to the driving. Secondly, the answer extends
the theory of adiabatic response of quantum systems undergoing
unitary evolution\cite{thouless83,berry93} to bipartite  quantum
systems consisting of a quantum  system and a quantum driving field
\cite{kibis10,dora12,trif12,gulasci15}. As a result, the presented
formalism opens a remarkable new area for response theory, where
condensed matter physics and quantum optics meet.

\section{formalism}
As a starting point, let us consider a generic two-band Hamiltonian,
\begin{eqnarray}\label{h0}
H_{0}(\vec{k})=\vec{d}(\vec{k})\cdot\vec{\sigma}+\epsilon(\vec{k})\cdot\textbf{I},
\end{eqnarray}
where $\textbf{I}$ is the $2\times 2$ identity matrix,
$\vec{\sigma}=(\sigma_x,\sigma_y,\sigma_z)$ are Pauli matrices,
$\epsilon(\vec{k})$ and $\vec{d}(\vec{k})$ depend on the materials
under study  and determine its band structure.  The two bands may
describe different physical degrees of freedom. If they are the
components of a spin-1/2 electron, $\vec{d}(\vec{k})$ stands for the
spin-orbit coupling. If they denote the orbital degrees of freedoms,
then $\vec{d}(\vec{k})$ represents the hybridization between bands.
The discussion below is completely independent of the physical
interpretation of the Hamiltonian Eq. (\ref{h0}), and leads to a
general formalism regarding the two-band system.

In the next section, we will specify $\epsilon(\vec{k})$ and
$\vec{d}(\vec{k})$ to examine the response of a concrete quantum
system to a quantum driving field. In the presence of a
electromagnetic field represented by vector potential $\vec{A}$ of
frequency $\omega$, by changing the crystal momentum,
$\vec{k}\rightarrow(\vec{k}-\frac{e}{\hbar}\vec{A}),$  we can still
use the two-band model to describe the system  in the field. In the
weak field limit, we may expend the Hamiltonian up to the first
order in $\vec{A}$,
\begin{eqnarray}\label{h}
H=H_0(\vec{k}-\frac{e}{\hbar}\vec{A}) \simeq
H_{0}(\vec{k})-\frac{e}{\hbar}\sum_{j=x,y,z}(\nabla d_{j}\cdot
\vec{A})\cdot \sigma_{j}.
\end{eqnarray}
In the Hilbert space spanned by the eigenstates of $\sigma_z$,
satisfying  $\sigma_z|\Uparrow\rangle=+|\Uparrow\rangle,$
$\sigma_z|\Downarrow\rangle=-|\Downarrow\rangle,$ the eigenvalues
and the corresponding eigenstates of $H_{0}(\vec{k})$ take,
$\varepsilon_{\pm}=\epsilon(\vec{k})\pm |\vec{d}|$ and $
|\varepsilon_{+}\rangle= \cos \frac{\theta}{2}
e^{-i\phi}|\Uparrow\rangle +\sin \frac{\theta}{2}
|\Downarrow\rangle$, $|\varepsilon_{-}\rangle= \sin \frac{\theta}{2}
e^{-i\phi}|\Uparrow\rangle -\cos \frac{\theta}{2}
|\Downarrow\rangle$. Here,
$|\vec{d}|=\sqrt{d_{x}^{2}+d_{y}^{2}+d_{z}^{2}}$,
$\cos\theta=\frac{d_{z}}{|\vec{d}|}$, and
$\tan\phi=\frac{d_{y}}{d_{x}}.$

Taking the field to be in the $x-$direction, $\vec{A}=(A_x,0,0),$
and decomposing  the field in a mean amplitude $\bar{E}$  and a
quantum part, $\delta_E(a^{\dagger}+a)$, i.e.,
\begin{equation}\label{vp}
A_{x}=E_{x}t=\bar{E}t+\delta_E(a^{\dagger}+a)t,
\end{equation}
we write the Hamiltonian as,
\begin{eqnarray}\label{6}
H&=&|\vec d|\tau_{z}+(g_c|\varepsilon_+\rangle\langle
\varepsilon_-|e^{i\omega t}+h.c.)\nonumber\\
&+&\left [g_{q}|\varepsilon_+\rangle\langle
\varepsilon_-|(a^{\dag}e^{-i\omega t}+ae^{i\omega t})+h.c.\right ].
\end{eqnarray}
Here $g_c\equiv i e \bar{E}\langle \varepsilon_{+}|
\frac{\partial\varepsilon_{-}}{\partial k_{x}}\rangle$, $g_q\equiv
 i e \delta_E\langle \varepsilon_{+}|
\frac{\partial\varepsilon_{-}}{\partial k_{x}}\rangle.$
$\tau_{+}\equiv|\varepsilon_{+}\rangle\langle\varepsilon_{-}|$,
$\tau_{-}\equiv\tau_+^{\dagger}$, and
$\tau_{z}\equiv|\varepsilon_{+}\rangle\langle\varepsilon_{+}|-|\varepsilon_{-}\rangle\langle\varepsilon_{-}|.$
$\bar{E}$ and $\delta_E$ are real, $a^{\dagger}$ and $a$ stands for
the creation and annihilation operator of the quantum part of the
field.

In terms of eigenstates of $H_c$ defined by $H_c\equiv |\vec
d|\tau_{z}+(g_c|\varepsilon_+\rangle\langle \varepsilon_-|e^{i\omega
t}+h.c.),$  the Hamiltonian can be rewritten as,
\begin{equation}\label{hc}
H=\sum_{j=+,-}E_j^{c}|E_j^c\rangle\langle E_j^c|+\hbar\omega
a^{\dagger}a+\eta(a^{\dagger}+a)|E_+^c\rangle\langle E_-^c|+h.c..
\end{equation}
Here,
$\eta=-g_q\cos^2\frac{\alpha_c}{2}e^{-i\beta_c}+g_q^*\sin^2\frac{\alpha_c}{2}e^{i\beta_c}$,
$E_{\pm}^c=\pm\sqrt{(|\vec{d}|-\frac{\hbar\omega}{2})^2+|g_c|^2}$,
$|E_+^c\rangle=\cos\frac{\alpha_c}{2}e^{i\beta_c}|\varepsilon_+\rangle+\sin\frac{\alpha_c}{2}|\varepsilon_-\rangle,$
 and $|E_-^c\rangle=\sin\frac{\alpha_c}{2}e^{i\beta_c}|\varepsilon_+\rangle-\cos\frac{\alpha_c}{2}|\varepsilon_-\rangle.$
$\cos\alpha_c=\frac{2|\vec{d}|-\hbar\omega}{\sqrt{(2|\vec{d}|-\hbar\omega)^2+4|g_c|^2}}$,
$\tan\beta_c=\Im (g_c)/\Re(g_c)$ with $\Im(...)$ and $\Re(...)$
denoting the imaginary and real part of $(...),$ respectively.

Under the rotating-wave approximation(RWA), the eigenstate and the
corresponding  eigenvalues take,
\begin{eqnarray}\label{8}
|E_{\pm}^q\rangle_{n}&=& \cos \frac{\alpha_q^\pm}{2}
e^{i\beta_q}|E_{+}^c\rangle\otimes|n\rangle +\sin
\frac{\alpha_q^\pm}{2} |E_{-}^c\rangle\otimes|n+1\rangle,
\end{eqnarray}
where
$\cos\alpha_q^+=\frac{\Delta}{\sqrt{\Delta^{2}+4|\eta|^2(n+1)}}$,
$\alpha_q^-=\alpha_q^+-\pi,$
$\tan\beta_q=\frac{\Im(\eta)}{\Re(\eta)}$, and
$\Delta=2E_+^c-\hbar\omega.$ $|n\rangle$ denotes a Fock state of the
field.  The results beyond the RWA will be given in Appendix. Using
the relation $v_y=\frac 1 \hbar \frac{\partial H(\vec{k})}{\partial
\vec{k}},$ we easily  find $v_y=\frac{\hbar
k_{y}}{m^{*}}+\sum_{j=x,y,z}\frac{\partial d_{j}}{\hbar\partial
k_{y}}\cdot \sigma_{j}.$  Then the $y-$component of the average
velocity in state $|E_-^q\rangle$ is given by,
\begin{eqnarray}\label{vy}
\bar{v}_y&=&_{n}\langle E_{-}^q
|v_y|E_{-}^q\rangle_{n}\nonumber\\
&=&\sin^{2}\frac{\alpha_q^-}{2} \langle
E_{+}^c|v_y|E_{+}^c\rangle+\cos^{2}\frac{\alpha_q^-}{2} \langle
E_{-}^c|v_y|E_{-}^c\rangle\nonumber\\
&=&-\cos\alpha_q^-\Re(\sin\alpha_c
e^{-i\beta_c}\langle\varepsilon_+|v_y|\varepsilon_-\rangle).
\end{eqnarray}
Consider the system under an external electric field $E_x\neq 0$
without magnetic field. The dc current density
$j(\bar{E},\delta_E,n)=j_y(\bar{E},\delta_E,n)$ can be then obtained
from the equation given  above by,
\begin{equation}\label{current}
j_y(\bar{E},\delta_E,n)=-e\int\frac{dk_x
dk_y}{(2\pi)^2}\bar{v}_y\big |_{\omega\rightarrow 0}.
\end{equation}
For the quantum part of the field, a linear response of the system
to the photon number in the field is then defined by,
\begin{equation}\label{sig_n}
\sigma_n=\frac{\partial j(\bar{E},\delta_E,n)}{\partial n}\big
|_{n\rightarrow 0}.
\end{equation}
After some straightforward algebra, and expanding $\sigma_n$ up to
the first order in $\bar{E}$, we have
\begin{widetext}
\begin{equation}\label{sig_n1}
\sigma_n=
\frac{e^2\bar{E}}{\hbar}\int\frac{idk_xdk_y}{(2\pi)^2}\frac{\partial
\cos\alpha_q^-}{\partial n}\bigg|_{n=0}\left[
\langle\frac{\partial\varepsilon_-}{\partial
k_x}|\frac{\partial\varepsilon_-}{\partial
k_y}\rangle-\langle\frac{\partial\varepsilon_-}{\partial
k_y}|\frac{\partial\varepsilon_-}{\partial k_x}\rangle\right].
\end{equation}
\end{widetext}
Noticing that the Berry curvature of the lower bare band
$|\varepsilon_-\rangle$ is defined by
$\Omega^-_{xy}(\vec{k})=i\left[
\langle\frac{\partial\varepsilon_-}{\partial
k_x}|\frac{\partial\varepsilon_-}{\partial
k_y}\rangle-\langle\frac{\partial\varepsilon_-}{\partial
k_y}|\frac{\partial\varepsilon_-}{\partial k_x}\rangle\right]$, we
find that $\sigma_n$ is simply the BZ integral of the Berry
curvature weighted by the factor $\frac{\partial
\cos\alpha_q^-}{\partial n}\bigg|_{n=0}$. Discussions on Eq.
(\ref{sig_n1}) are in order. Consider a limit of $\Delta\gg
4|\eta|^2(n+1)$, $\cos\alpha_q^-\sim
\frac{2|\eta|^2(n+1)}{\Delta^2}-1,$ then $\sigma_n\simeq
\frac{e^2}{h}\bar{E}\frac{2|\eta|^2}{\Delta^2}C_n.$  Here, we assume
$\frac{2|\eta|^2}{\Delta^2}$ independent of $\vec{k},$ and $C_n$
denotes the Chern number of band $|\varepsilon_-\rangle.$ This
suggests that $\sigma_n$ behaves like the conventional Hall
conductance. In fact, as will be seen below, the response of the
two-band system to the photon number in the field witnesses the
transition points of the system.

In addition, we may define a response of the topological insulator
to the mean amplitude of the field, taking the quantum part of the
field into account. Namely, define
\begin{eqnarray}
\sigma_q=\frac{\partial j(\bar{E},\delta_E,n)}{\partial
\bar{E}}\bigg |_{\bar{E}=0}
\end{eqnarray}
to characterize the response of the two-band system to the classical
part of the field. Simple algebra shows that,
$\sigma_q=\frac{e^2}{\hbar}\int\frac{dk_xdk_y}{(2\pi)^2}\cos\alpha_q^-\Omega^-_{xy}(\vec{k}).$
A limiting case for $\sigma_q$  is that
$\sigma_c=\sigma_q\big|_{\delta_E=0}=\frac{\partial
j(\bar{E},\delta_E,n)}{\partial \bar{E}}\bigg
|_{\bar{E}=\delta_E=0}$ quantifies the linear response of the
insulator to the mean amplitude $\bar{E}$ without quantum fields.
Clearly, with $\delta_E=0$ and  $\omega\rightarrow 0$, we have
$\eta=0$ and $\cos\alpha_q^-=1$. In this case, $\sin\alpha_c\simeq
\frac{g_c}{|\vec{d}|}$,  and $\sigma_c$ reduces to the well-known
result,
$$\sigma_c=\frac{e^2}{\hbar}\int\frac{dk_xdk_y}{(2\pi)^2}\Omega^-_{xy}(\vec{k}).$$
We should notice that $\sigma_c$ is exactly the conventional Hall
conductance, while $\sigma_q$ can be understood as the Hall
conductance under the influence of quantum fluctuations. In this
sense, we interpret $\sigma_q$ as the Hall conductance in quantized
fields, and $\sigma_n$ quantifies the response of the two-band
system to  photon number of the field. In the next section, we will
exemplify these responses with concrete examples.

\section{examples}

For an explicit discussion on the  Hall conductance, we first
consider the following choices of $\vec{d}(\vec{k})$, ${d_x} = \sin
{k_y},{d_y} = \sin {k_x},{d_z} = 2 - \cos {k_x} - \cos {k_y} - e_s$
\cite{qi06}.
\begin{figure}[h]
\centering
\includegraphics[width=1\columnwidth,height=0.8\columnwidth]{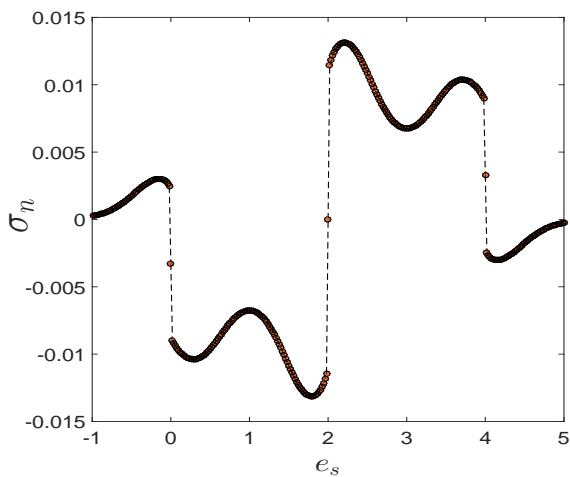}
\caption{(Color online) $\sigma_n$ (in units of $\frac{e^2}{h}$),
which quantifies the response of the system to the quantized part of
field, as a function of $e_s$ in a model with ${d_x} = \sin
{k_y},{d_y} = \sin {k_x},{d_z} = 2 - \cos {k_x} - \cos {k_y} - e_s$.
The other parameters chosen are $\delta_E=0.3$ meV/nm,
$\bar{E}$=0.1meV/nm.} \label{sigma_n1}
\end{figure}
Physically, this model can be interpreted  as a tight-binding model
describing a magnetic semiconductor with Rashba type spin-orbit
coupling, spin dependent effective mass and a uniform magnetization
on $z-$direction. It has been shown\cite{qi06} that $\sigma_c=1$ for
$0<e_s<2$; $\sigma_c=-1$ for $2< e_s<4$, and $\sigma_c=0$ for
$e_s<0$ and $e_s>4.$

\begin{figure}
\includegraphics[width=1\columnwidth,height=0.8\columnwidth]{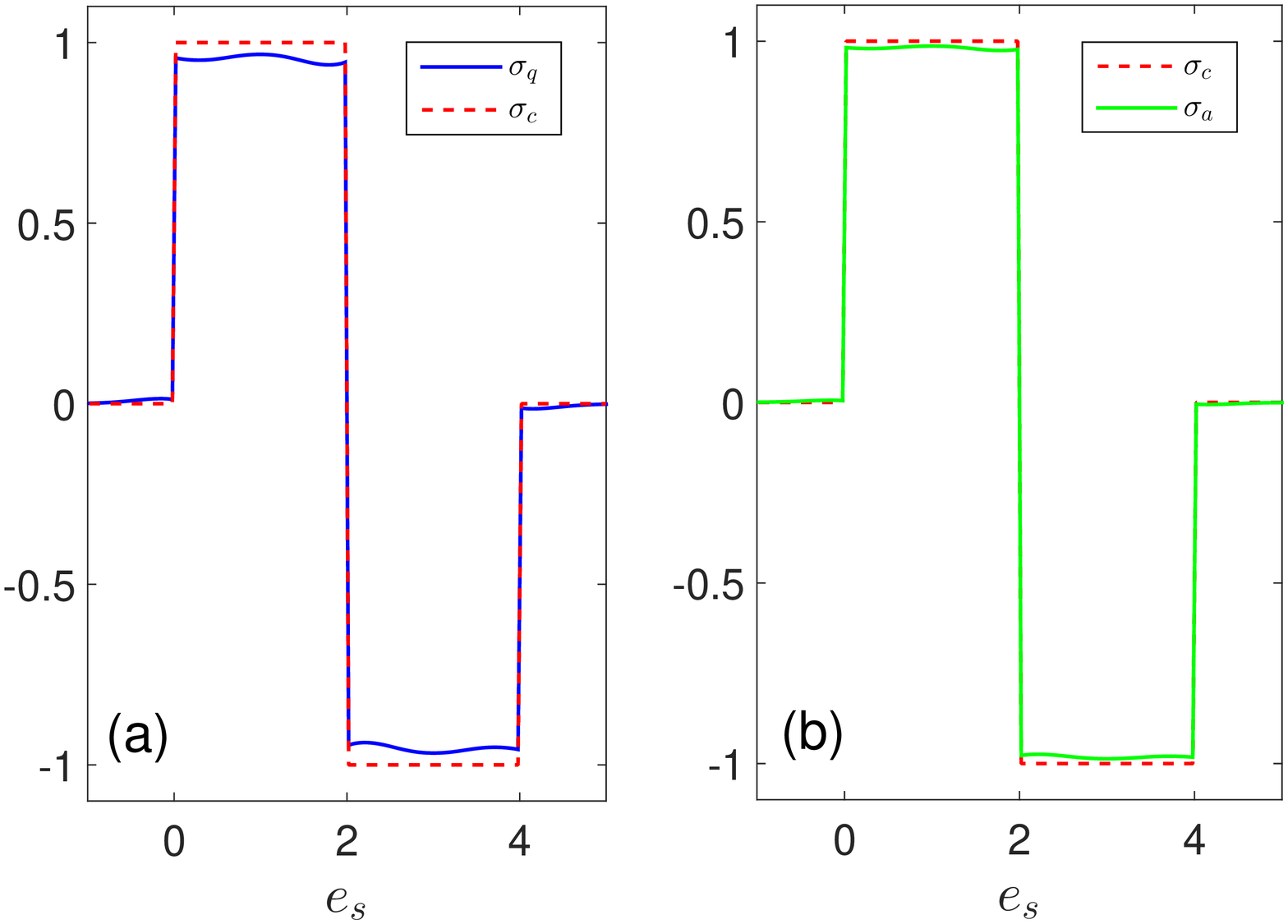}
\caption{ (color online)  (a) Hall conductance  $\sigma_q$(in units
of $e^2/h$) as a function of $e_s.$ $\delta_E=0.3$ meV/nm, $n=4$.
For comparison, the conventional Hall conductance (red-dashed line)
is also shown.  (b) Averaged Hall conductance $\sigma_a$ as a
function of $e_s$. $\sigma_a$ is defined as $ {\sigma_a} =
\frac{1}{N}\sum_{j=1}^N\sigma_q(\delta_E^j),$ where $\delta_E^j$
denotes the $j-$th random value of $\delta_E$ from  $[-0.3,0.3].$
Here $N=50$. $\vec{d}(\vec{k})$ is the same as in Fig.
\ref{sigma_n1}. } \label{averf1}
\end{figure}

With respect to the photon number $n$, the Hall conductance
$\sigma_n$  defined in Eq. (\ref{sig_n})  is plotted as a function
of $e_s$ in Fig. \ref{sigma_n1}. We find that the phase transition
points, i.e., $e_s=0, 2, 4 $ remain unchanged. In contrast with the
well known Hall conductance $\sigma_c$ shown in Fig. \ref{averf1}
(red dashed lines), $\sigma_n$ is not a constant in regions,
$0<e_s<2$, $2< e_s<4$,  $e_s<0$ and $e_s>4.$ This results from the
weight $\frac{\partial\cos\alpha_q^-}{\partial n}$ in the integral
of Eq. (\ref{sig_n1}).  Physically, the weight plays the role of
distribution function, which is not a constant and depends on $k_x$,
$k_y$ and $e_s$ in this model. Fig. \ref{averf1} shows $\sigma_c$,
$\sigma_a$ and $\sigma_q$ as a function of $e_s$, where $\sigma_a$
is defined as $\sigma_a=\frac 1 N \sum_{j=1}^N\sigma_q(\delta_E^j)$.
$\delta_E^j$ denotes the $j-$th value of $\delta_E$ randomly chosen
from $[-0.3, 0.3],$ that is, $\sigma_a$ is defined as an average
over $\delta_E$ chosen randomly in interval $[-0.3, 0.3].$  Two
observations can be made. (1) Quantum fluctuations suppress  the
Hall conductance $\sigma_c$, but they do not change the phase
transition points; (2) $\sigma_a$ is very close to $\sigma_c$,
suggesting that the quantum fields (fluctuations of the classical
field)  have small effect on the Hall conductance on average.

The second example we will take to illustrate the conductances  is a
two-dimensional lattice in a magnetic field \cite{kohmoto89}. The
tight-binding Hamiltonian for such a lattice takes,
$H=-t_a\sum\limits_{\left\langle {i,j} \right\rangle } {_x}
c_j^\dagger {c_i}{e^{i{\theta _{ij}}}}-t_b\sum\limits_{\left\langle
{i,j} \right\rangle } {_y} c_j^\dagger {c_i}{e^{i{\theta _{ij}}}},$
where $c_j$ is the usual fermion operator on the lattice.   The
phase $\theta_{ij}=-\theta_{ji}$ represents the magnetic flux
through the lattice.  When $t_b=0$, the single band  is doubly
degenerate. The term with $t_b$ in the Hamiltonian gives the
coupling between the two branches of the dispersion.  Consider two
branches which are coupled by $|l|-$th order perturbation, the gaps
open and the size of the gap due to this coupling is the order of
$t_{b}^{|l|}$. The effective Hamiltonian then take the form of Eq.
(\ref{h0}) with ${d_x} = \delta \cos {k_y},{d_y} = \delta \sin
{k_y},{d_z} = 2{t_a}\cos ({k_x} + 2\pi mp/q)$, where $p,q$ are
integers, $\delta$ is proportional to (is the order of) $t_b^{\left|
l \right|}.$

\begin{figure}[h]
\centering
\includegraphics[width=1\columnwidth,height=0.8\columnwidth]{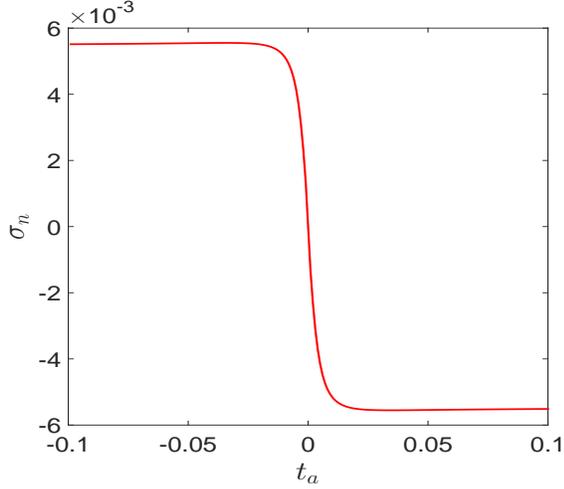}
\caption{(Color online) $\sigma_n$ (in units of $\frac{e^2}{h}$) as
a function of $t_a$. In this model,  ${d_x} = \delta \cos
{k_y},{d_y} = \delta \sin {k_y},{d_z} = 2{t_a}\cos ({k_x} + 2\pi
mp/q)$. The other parameters are $\delta_E=0.3$ meV/nm,
$\bar{E}$=0.1meV/nm, $m = 1, p = 1, q = 4$, $\delta=0.01$ meV.}
\label{sigma_n2}
\end{figure}
\begin{figure}
\includegraphics[width=1\columnwidth,height=0.8\columnwidth]{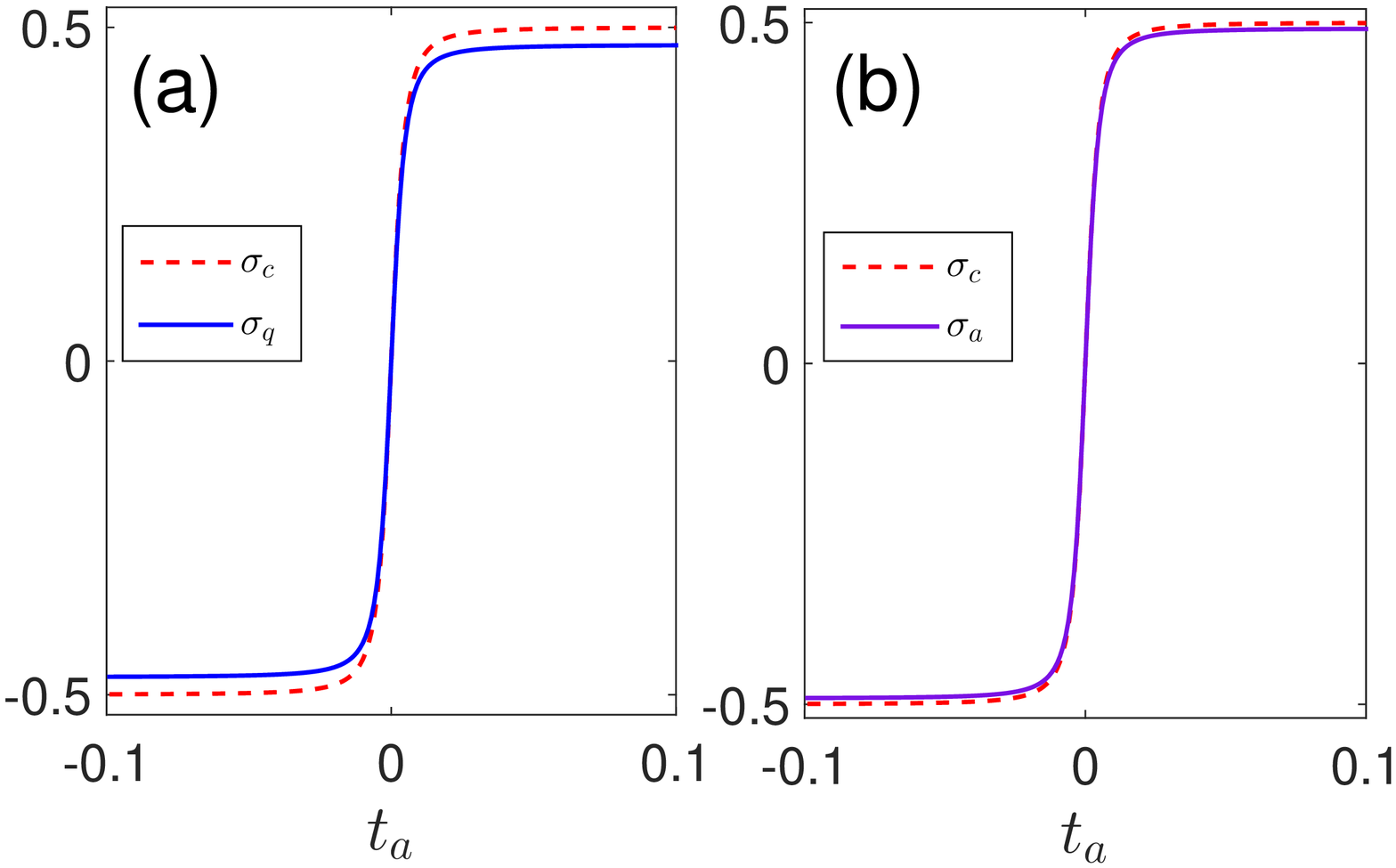}
\caption{ (color online) (a) Hall conductance $\sigma_q$  as a
function of $t_a$. For comparison, the conventional Hall conductance
$\sigma_c$ is also plotted. The model is ${d_x} = \delta \cos
{k_y},{d_y} = \delta \sin {k_y},{d_z} = 2{t_a}\cos ({k_x} + 2\pi
mp/q)$. The other parameters chosen are $\delta_E=0.3$ meV/nm. (b)
Averaged Hall conductance $\sigma_a$ versus $t_a$. $\sigma_a$ is
defined in the same way as in Fig. \ref{averf1}. $\delta_E$ is
randomly chosen from [-0.3,0.3] meV/nm for 50 times. The other
parameters chosen  for both (a) and (b) are $m = 1, p = 1, q = 4,$
$\delta=0.01$ meV, $n=4.$ All conductances are plotted in units of
$\frac{e^2}{h}$.} \label{averf2}
\end{figure}

From Fig. \ref{sigma_n2}, we observe that $\sigma_n$ is very small,
but it can witness  the phase transition points. Fig. \ref{averf2}
shows the conventional Hall conductance $\sigma_c$, the Hall
conductance  $\sigma_q$ subject to the quantized field, and the
averaged Hall conductance $\sigma_a$ as a function of $t_a.$ We find
that the transition points remain unchanged, but the Hall
conductance is slightly changed. The features observed from Fig.
\ref{sigma_n2} and Fig. \ref{averf2} support the conclusions made in
Fig. \ref{sigma_n1} and Fig. \ref{averf1}. These observations
suggests that the quantum Hall effect can be taken as a method to
determine the fine structure constant even in the presence of
quantum fluctuations.

It is worth noticing that all hall conductance including $\sigma_q,$
$\sigma_c$ and $\sigma_a$ are zero when $\bar E =0,$ since in this
case,
$$\bar{v}_y=\sin^2\frac{\alpha_{q0}}{2}\langle\varepsilon_+|v_y|\varepsilon_+\rangle
+\cos^2\frac{\alpha_{q0}}{2}\langle\varepsilon_-|v_y|\varepsilon_-\rangle=0.$$
Here $\alpha_{q0}=\alpha_{q}^-(\bar E=0).$ In other words, a
quantized field can not induce  current in the system. This feature
is reminiscent of the which-way experiment\cite{durt98,buks98} that
an attempt to gain information about the path taken by the particle
inevitably reduces the visibility of the interference pattern. Here
the quantum field can record the information of the path, while the
classical field can not. Indeed, observing Eq.(\ref{vy}), we find
that the current induced by the external field is very similar to
the interference patten in the which-way experiment,  where
$|\varepsilon_+\rangle$ and $|\varepsilon_-\rangle$ play the role of
the two paths.

\begin{figure}[h]
\centering
\includegraphics[width=1\columnwidth,height=0.8\columnwidth]{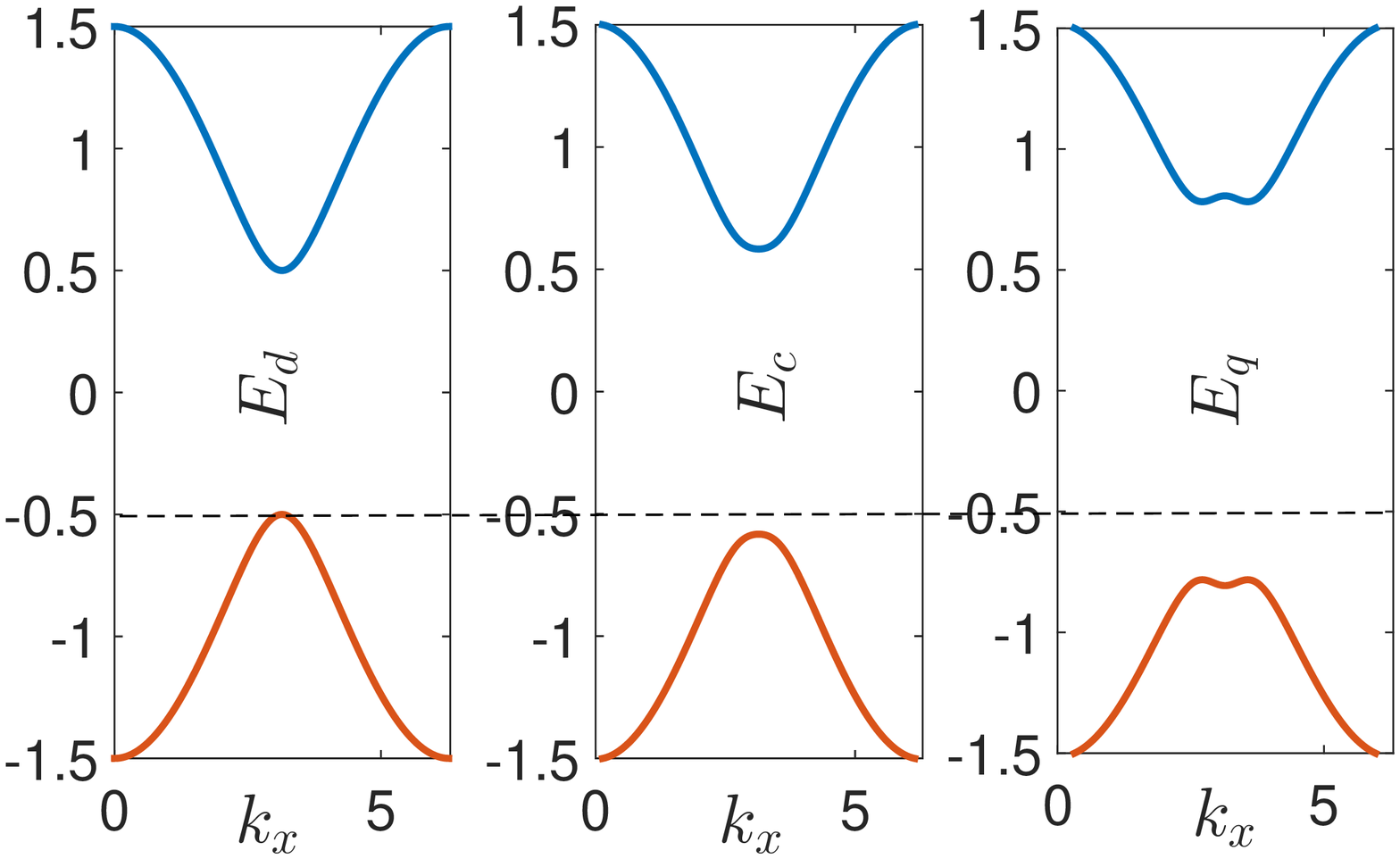}
\caption{(Color online) Energy spectrum (in units of meV) of the
system in the first example. The other parameters are $e_s=1.5$meV.
$\delta_E=\bar{E}=0.3$ meV/nm,  $n=0, \omega=0.$ } \label{ef1}
\end{figure}

Consider the  case without photon in the field and neglect the
vacuum effect, i.e., $n=0$ and $\omega=0$, the change in Hall
conductances (with respect to the conventional Hall conductance) can
be understood as a consequence of band mixing caused by the quantum
field, since the bulk band gaps remain open, see Fig. \ref{ef1}. In
Fig. \ref{ef1}, we plot the energy spectrum of the system in the
first example. $E_d$ denotes the spectrum of the system $H_0$
without external fields, $E_c$ stands for the spectrum of the system
in the external field with $\delta_E=0$, and $E_q$ is the spectrum
with $\delta_E\neq 0$.  The interactions parameterized by $\bar E$
and $\delta_E$ enlarge the band gaps. So, the topological nature of
the system remains unchanged.

\begin{figure}[h]
\centering
\includegraphics[width=1\columnwidth,height=1\columnwidth]{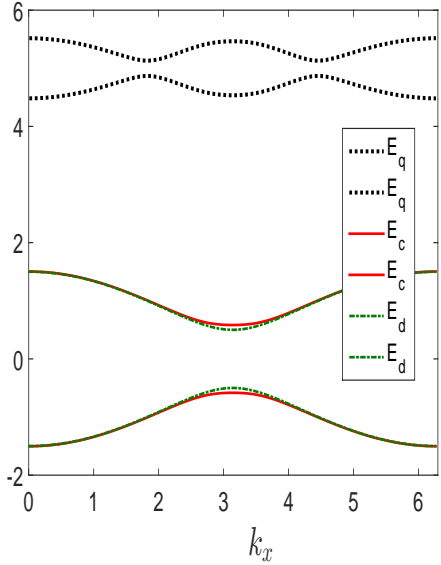}
\caption{(Color online) The same as Fig. \ref{ef1}, but with
$\omega=2, n=1.$} \label{ef2}
\end{figure}

The result changes when $n\neq 0$ and $\omega\neq 0.$ The quantized
field (or the photon field) can change the topology of the system,
see Fig.\ref{ef2}. It is possible to switch between different
topological phases by changing the photon number and the frequency,
which may induces more avoid-crossing points  as depicted in
Fig.\ref{ef2}. This observation is confirmed by an ac conductance
$\sigma_q(\omega)$, which is defined in the same way as $\sigma_q$
but without the limitation of $\omega\rightarrow 0$ in Eq.
(\ref{current}).

\begin{figure}[h]
\centering
\includegraphics[width=1\columnwidth,height=0.8\columnwidth]{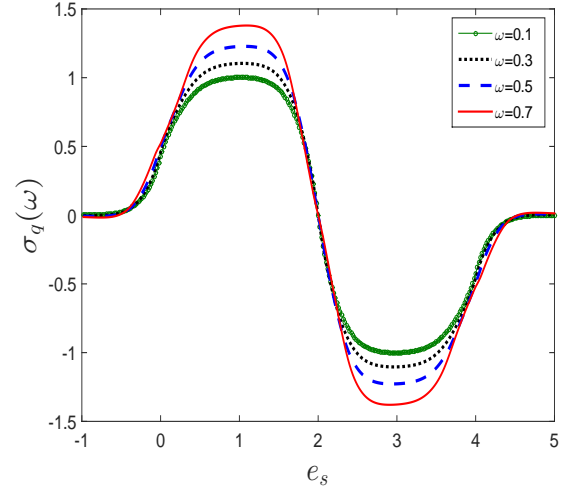}
\caption{(Color online) Hall conductance $\sigma_q(\omega)$ ( in
units of $\frac{e^2}{h}$) against $e_s$ at various  frequencies  of
the external field. In this plot, ${d_x} = \sin {k_y},{d_y} = \sin
{k_x},{d_z} = 2 - \cos {k_x} - \cos {k_y} - e_s$. The other
parameters are $\delta_E=0.3$ meV/nm, $E_x=0.1$ meV/nm, $n=4.$}
\label{res_qome}
\end{figure}

To show the ac conductance of the system in the first example, we
calculate  $\sigma_q(\omega)$ as a function of $e_s$ at various
applied electric field frequencies $\omega$. The numerical results
are shown in Fig. \ref{res_qome}. The difference between
$\sigma_q(\omega)$ with various $\omega$ arises because the photon
field may induce  more avoid-crossing points,  which is depicted in
Fig.\ref{ef2}, lines for $E_q.$

\section{discussion and conclusion}
The two-band model may describe topological insulators, which is
realized by using either condensed matter\cite{lindner11} or cold
atoms settings\cite{jiang11}. The single mode field enters the
system via a vector potential. The single photon mode is realized in
a quantum LC circuit\cite{hepp73} or is selected from a ladder of
cavity models by placing a dispersive element into the caivty, of
which the reflective index is wave-vector dependent. Tuning
frequency $\omega$ and the coupling of the field with ITs is
possible by changing the dielectric constant. The Hall conductance
(equivalently the Chern number) can be probed through a Thouless
type\cite{ye11}. The photon number may be tuned  by a real-time
quantum feedback procedure that generates  on demand and stabilizes
photon number states by reversing the effects of decoherence-induced
quantum jumps\cite{sayrin11}. Alternatively, the photon number may
be tuned via changing the coupling constant, since the square root
of the photon number $\sqrt{n+1}$ always appears  with the coupling
constant $\eta.$

In summary, we have introduced the response of a two-band system to
a quantized single-mode field. Three types of Hall conductance are
introduced to quantify this response. Two examples are presented to
exemplify the theory. Physics behind the findings is revealed and
discussed.

\section*{ACKNOWLEDGMENTS}
This work is supported by National Natural Science Foundation of
China (NSFC) under Grants No. 11175032, and No. 61475033.

\appendix
\section{The result beyond the RWA}
In this APPENDIX, we will present discussions on the results beyond
the Rotating-wave approximation (RWA). We start with the Hamiltonian
in the maintext,
\begin{equation}
H=\sum_{j=\pm}E_j^c|E_j^c\rangle\langle E_j^c|+\hbar\omega
a^{\dagger}a+\eta(a^{\dagger}+a)|E_+^c\rangle\langle E_-^c|+h.c.
\end{equation}
Notations are the same as in the maintext. To solve this
Hamiltonian, we transform $H$ into an effective Hamiltonian,
\begin{equation}
H_{eff}=e^sHe^{-s}=\sum_{j=\pm}E_j^c|E_j^c\rangle\langle
E_j^c|+\hbar\omega a^{\dagger}a+g^\prime a|E_+^c\rangle\langle
E_-^c|+h.c.
\end{equation}
Here, $s=\frac{\Re({\eta})}{2E_+^c+\hbar\omega}\tau_x(a^\dagger-a)
-\frac{\Im({\eta})}{2E_+^c+\hbar\omega}\tau_y(a^\dagger-a),$ and
\begin{equation}
g^{\prime}=\frac{4E_+^c}{2E_+^c+\hbar\omega}\eta,
\end{equation}
The eigenstates of the effective Hamiltonian are then,
$$|E_{+}^q\rangle_{n}^\prime= \cos \frac{\alpha_q^\prime}{2}
e^{i\beta_q^\prime}|E_{+}^c\rangle\otimes|n\rangle +\sin
\frac{\alpha_q^\prime}{2} |E_{-}^c\rangle\otimes|n+1\rangle,$$ and
$$|E_{-}^q\rangle_{n}^\prime= \sin \frac{\alpha_q^\prime}{2}
e^{i\beta_q^\prime}|E_{+}^c\rangle\otimes|n\rangle -\cos
\frac{\alpha_q^\prime}{2} |E_{-}^c\rangle\otimes|n+1\rangle,$$ with
$\alpha_q^{\prime}$ being defined by,
$$\cos\alpha_q^\prime=\frac{(2E_+^c-\hbar\omega)}{\sqrt{\Delta^2+4|g^\prime|^2(n+1)}},$$
and $\Delta=(2E_+^c-\hbar\omega).$ The corresponding eigenenergies
are denoted by $E_+^{q\prime}(n)$ and $E_-^{q\prime}(n)$,
respectively. Assuming band $|E_{-}^q\rangle_{n}^\prime$ is filled,
we may calculate the current and the Hall conductance discussed
above. Obviously, the Hall conductance takes the same formula except
$\alpha_q^{\prime}.$ The difference between $\alpha_q^{\prime}$ and
$\alpha_q$ originates from the coupling constant
$g^{\prime}=\frac{4E_+^c}{2E_+^c+\hbar\omega}\eta.$ For points
$\{\vec{k}\}$ satisfying (resonant condition) $2E_+^c=\hbar\omega,$
we have $g^{\prime}=\eta,$ i.e., no difference $g^{\prime}$ and
$\eta$ at these resonant points. However, for the off-resonant
points, $g^{\prime}$ and $\eta$ might be very different, which can
lead to different topological phases.

\end{document}